\documentclass[twocolumn,amsmath,amssymb,floatfix,prd]{revtex4}

\usepackage{bm}
\usepackage{amsmath}
\usepackage{epsf}

\newcommand{\beq}{\begin{equation}}

\newcommand{\eeq}{\end{equation}}
\newcommand{\beqa}{\begin{eqnarray}}
\newcommand{\eeqa}{\end{eqnarray}}

\def\simlt{\lesssim}

\newcommand{\Hinf}{H_{\rm inf}}

\newcommand{\mnras}{Mon. Not. R. Astron. Soc.}

\begin{document}

\title{A Low CMB Quadrupole from Dark Energy Isocurvature Perturbations}
\author{Christopher Gordon$^{1}$ and Wayne Hu$^{1,2}$}
\affiliation{
$^{1}$Kavli Institute for Cosmological Physics
and Enrico Fermi Institute, 
University of Chicago, Chicago IL 60637 \\
$^{2}$Department of Astronomy and Astrophysics,
University of Chicago, Chicago IL 60637 
}

\begin{abstract}
\baselineskip 11pt
We explicate the origin of the temperature quadrupole in the adiabatic dark energy
model and explore the mechanism by which scale invariant isocurvature dark energy
perturbations
can lead to its sharp suppression.  The model requires anticorrelated curvature and
isocurvature fluctuations and  is favored by the WMAP
data at about the 95\% confidence level in a flat scale invariant model.  
In an inflationary context, the anticorrelation
may be established if the curvature fluctuations originate from a variable decay rate of
the inflaton; such models however tend to overpredict gravitational waves.   This isocurvature
model can in the future be distinguished from alternatives involving a reduction in
large scale power or modifications to the sound speed of the dark energy
through the polarization and its  cross correlation with the temperature.  
The isocurvature model retains
the same polarization fluctuations as its adiabatic counterpart but reduces
the correlated temperature fluctuations.
We present a pedagogical discussion of dark energy fluctuations in a quintessence and
k-essence context in the Appendix.
\end{abstract}
\maketitle

\section{Introduction}

The first data release of the WMAP data \cite{Ben03} was on the whole
in spectacular agreement with the $\Lambda$CDM model
\cite{Spe03,Pei03}.  However, as originally discovered by COBE,
the measured value of the quadrupole
seems low compared to the model prediction. Originally it was
estimated that the probability of measuring such a low or lower
quadrupole was 0.7\% assuming that the $\Lambda$CDM model was correct
\cite{Spe03}. However, subsequent studies with more detailed methods
of dealing with foregrounds have estimated this probability to be
closer to 4\% \cite{Efs03,Efs04,TegOliHam03,SloSelMak04,SloSel04}. The
alignment of the quadrupole and octopole
\cite{deOTegZalHam03,SchStaHutCop04,SloSel04,BiePawGorBan04} and
various asymmetries in the data
\cite{EriBanGorLil04,EriHanBanGorLil03,HanCabMarVit04,HanBanGor04,PruUzaBernBrun04,HanBalBanGor04}
have also being considered.  

This probability of such a low or lower quadrupole would not be
 particularly anomalous if the quadrupole were treated on par with
all other multipoles.  Such
points would be expected to occur just by chance in such a large data set.
In fact there are several other equally
or more anomalous multipoles.  However the quadrupole is particularly
intriguing in that it represents the largest observable angular scale and so may be
a good probe of new physical effects. In particular, it 
is the 
multipole whose power has the most  significant contribution from length scales
that are on and even above the horizon at dark energy domination.

Various explanations for the low quadrupole have been proposed. They
include a cut off in the primordial power spectrum
\cite{BriLewWelEfs03,Efs03a,Lin03,LasDor03,ConPelKofLin03,CliCrotLes03,PiaFenZha03,TsuSinMaa03,TsuMaaBran03,BasFreMers03,FenZha03,PiaTsuZha03,LigMatMusRio04,EnqSlo04},
a small universe
\cite{Spe03,LumWeeRiazLehUza03,WeeLumRia03,AurLusSteThe04,PhilKog04}
and perturbations in the dark energy
\cite{MorTak03,DedCalSte03,WelLew03,BeaDor04,AbrFinPer04}. 
The cut-off and small
Universe models work by reducing the Sachs Wolfe effect in the
quadrupole.  The dark energy perturbation models work by modifying the
Integrated Sachs Wolfe effect from the dark energy.
In this paper we focus on the latter class and in particular a generalization of 
the correlated isocurvature model introduced by \cite{MorTak03}.

We begin in \S \ref{sec:transfer} with a general discussion of the origin
of the temperature quadrupole in the adiabatic model and explore its relationship
to the low multipole polarization.   In \S \ref{sec:isocurvaturemodel}, we show how
the properties of the adiabatic quadrupole point to a specific class of isocurvature
models that can cancel the Sachs-Wolfe contributions to the quadrupole.
We compare and contrast this model with alternate solutions and show
that the polarization will be in the future a useful discriminator.
In \S \ref{sec:likelihood} we assess the likelihood of substantial isocurvature
perturbations in light of the WMAP data.  We discuss an inflationary context for
such perturbations in \S \ref{sec:inflation} but show that in the simplest models
gravitational waves are over-predicted.  We also include a pedagogical Appendix
on dark energy perturbations in a quintessence and k-essence context.

\section{Quadrupole Transfer Function}
\label{sec:transfer}

In an adiabatic model with dark energy, the CMB temperature quadrupole
receives its contributions from two distinct effects: the (ordinary) Sachs-Wolfe (SW)
effect from temperature and metric fluctuations near recombination and the
Integrated Sachs-Wolfe (ISW) effect from changes in the metric fluctuations due to
the dark energy.  These effects are quantified by the CMB temperature transfer function.

\begin{figure}[tb]
\centerline{\epsfxsize=3.4in\epsffile{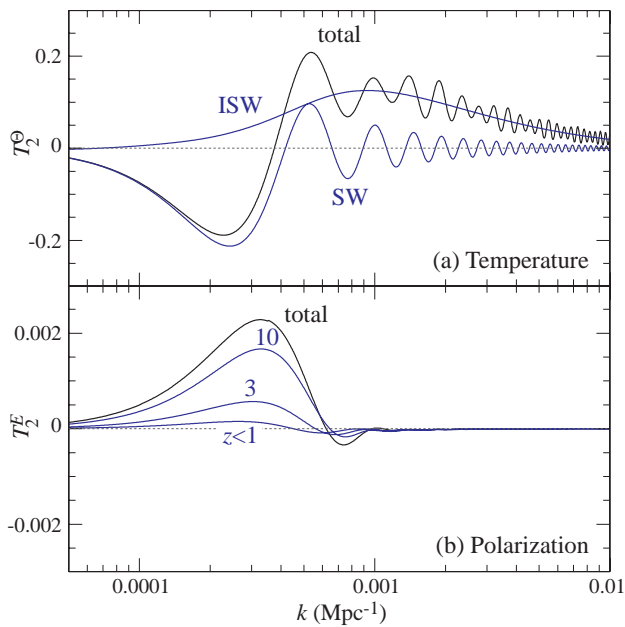}}
\caption{\footnotesize
The 
quadrupole transfer function in the fiducial adiabatic model (see text).  The temperature quadrupole
receives Sachs-Wolfe (SW) contributions peaking around
$k=0.0002$ Mpc$^{-1}$ and Integrated Sachs-Wolfe (ISW) contributions peaking
around
$k=0.001$ Mpc$^{-1}$ but extending to $k\sim 0.01$ Mpc$^{-1}$.  
The ISW effect arises from the
dark energy dominated regime $z \simlt 1$.  The polarization arises through rescattering 
of quadrupole anisotropies at $z > 1$ and hence
reflects the SW contributions.  The cross correlation is proportional
to the product of the transfer functions. }
\label{fig:quadadi}
\end{figure}

Let us define the two dimensional CMB transfer functions
as the mapping between the power in the initial curvature fluctuations $\zeta_{i}$
 in the comoving gauge
[see Appendix, Eqn.~(\ref{eqn:bardeencurvature})] 
\begin{equation}
\langle \zeta_i({\bf k})\zeta_i({\bf k}') \rangle =
(2\pi)^3 \delta({\bf k}-{\bf k}') {2\pi^2 \over k^3} \Delta_{\zeta_i}^2(k)\,
\end{equation}
and the angular space power spectra 
\begin{equation}
{\ell (\ell+1) C_\ell^{XX'} \over 2\pi} = \int {dk \over k}
T_\ell^X(k,\eta_0) T_\ell^{X'}(k,\eta_0) \Delta_{\zeta_i}^2(k)\,,
\label{eqn:cl}
\end{equation}
where $X,X' \in \Theta,E$ the temperature fluctuation and $E$-mode polarization
respectively.  

\begin{figure}[tb]
\centerline{\epsfxsize=3.4in\epsffile{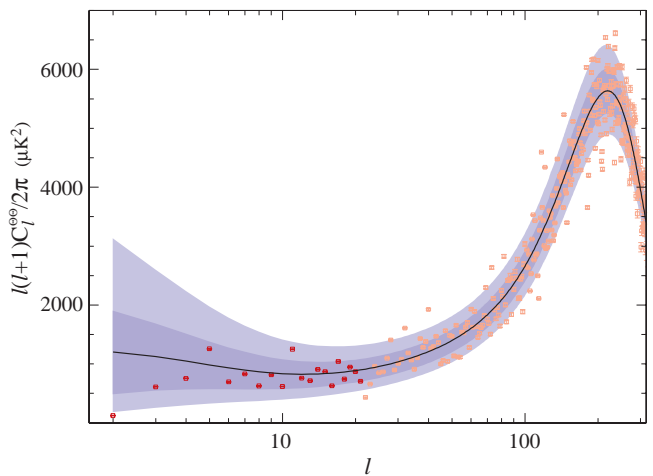}}
\caption{\footnotesize
Temperature power spectrum in the fiducial model compared with the data from the first
year WMAP release \cite{Ver03}.  Bands are 68\% and 95\% 
cosmic variance confidence regions (see text).}
\label{fig:cladi}
\end{figure}

We employ a comoving gauge Boltzmann hierarchy code \cite{HuOka03} 
for numerical solutions of the transfer function.   These are shown for the temperature
and polarization quadrupole in Fig.~\ref{fig:quadadi}.   Here we have chosen 
fiducial values for the cosmological parameters that are near the maximum likelihood
model from WMAP: a dark energy density relative to critical of $\Omega_{Q}=0.73$,
non-relativistic matter density $\Omega_m h^2 = 0.14$, baryon density $\Omega_b h^2=0.024$,
optical depth to reionization $\tau=0.17$, dark energy equation of state $w_Q = p_Q/\rho_Q = -1$ in a spatially flat universe.
The resulting temperature
power spectrum is shown in Fig.~\ref{fig:cladi}  compared with the WMAP data for a scale invariant spectrum of initial perturbations
\begin{equation}
\Delta_{\zeta_i}^2 = \delta_{\zeta_i}^2 \left( {k \over 0.05 {\rm Mpc}^{-1} }\right)^{n-1}
\label{eqn:fiducialspect}
\end{equation}
where $\delta_{\zeta_i} = 5.07 \times 10^{-5}$ and the tilt $n=1$.

Notice that the temperature contributions from the SW and ISW effects are 
comparable in magnitude and
well-separated in scale.  Hence they add nearly in quadrature in Eqn.~(\ref{eqn:cl}).
In the polarization, the quadrupole comes mainly from the SW effect as can be
seen from its dependence on the redshift of reionization
or equivalently the cumulative contributions from $z<z_{\rm max}$ (see 
Fig.~\ref{fig:quadadi}b).  The polarization
quadrupole gets nearly no contributions from $z<1$ when the dark energy
dominates.   These properties are the key to understanding how to construct a model
with certain desired properties at low temperature and polarization multipoles. 

Since the quadrupoles are dominated by large scale or low-$k$ fluctuations, it is useful
to examine the origin of these properties with a low-$k$ approximation to the transfer
functions. In this limit, the transfer function is determined by the Newtonian temperature monopole $\Theta$, 
gravitational potential $\Psi$ and curvature
fluctuation $\Phi$ [see Eqn.~(\ref{eqn:phi}), (\ref{eqn:psi}) for the correspondence
to the comoving gauge]
\begin{eqnarray}
T_\ell^\Theta(k,\eta_0) &=& {\sqrt{2 \ell(\ell+1)} \over \zeta_i} 
\Big[  {({\Theta_* +\Psi_*}) j_\ell(kD_*)}  \nonumber\\
&& + \int_{\eta_*}^{\eta_0}d\eta
 (\dot \Psi -\dot \Phi)j_\ell(kD) \Big] \nonumber\\
 & \approx & - {\sqrt{2 \ell(\ell+1)} \over \zeta_i} 
\Big[  {1 \over 5}\zeta_{i}  j_\ell(kD_{\rm m}) \nonumber\\
&& + 2\int_{\eta_{\rm m}}^{\eta_0}d\eta \,
 \dot \Phi \, j_\ell(kD) \Big] \,,
 \label{eqn:analytictransfer}
\end{eqnarray}
where the subscripts denote evaluation at recombination for $``*"$, 
$``0"$ for the present and ``m" for some arbitrary time well after radiation
domination but well before dark energy domination and overdots represent 
derivatives with respect to conformal time $\eta=\int dt/a$.
Here we have assumed a spatially flat universe where the comoving distance
$D =\eta_0 -\eta$.  Since $\Phi \approx -\Psi$ when the anisotropic stress
is negligible [see Eqn.~(\ref{eqn:psi})], we loosely refer to either as the gravitational
potential.

\begin{figure}[tb]
\centerline{\epsfxsize=3.2in\epsffile{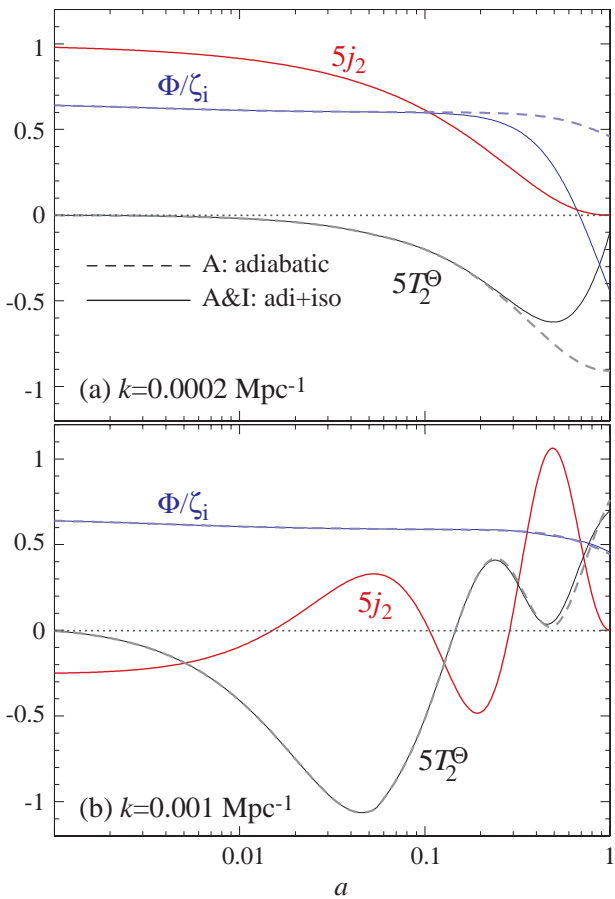}}
\caption{\footnotesize
Time evolution of gravitational potential $\Phi$ and quadrupole $T_2^\Theta$ in the
fiducial adiabatic model compared an adiabatic plus isocurvature model with
$S_{i} \equiv \delta_{Qi}/\zeta_{i}=-15$
(A\&I; see \S \ref{sec:isocurvaturemodel}).  
(a) $k=0.0002$ Mpc$^{-1}$ SW dominated mode.  In the
adiabatic model, the decay of the potential at $z\simlt 1$ has little effect on the quadrupole
due to the inefficiency in the transfer represented by $j_2$.  A much larger change of
$\Delta \Phi \sim -\zeta_i$ from the isocurvature perturbation can cancel the SW
quadrupole.  (b) $k=0.001$ Mpc$^{-1}$ ISW dominated mode.  The adiabatic potential
decay transfers efficiently onto the quadrupole moment.   The additional isocurvature
perturbations decay by dark energy domination and have little effect (see Fig.~\ref{fig:qevol}).}
\label{fig:time}
\end{figure}

The two terms in the second line are the SW and ISW effects respectively.
This approximation accounts for the small evolution in
the gravitational potential between recombination and full matter domination,
sometimes called the ``early" ISW effect.   Since $D_{\rm m} \approx D_*$, this
effect adds coherently with those at recombination.    In the fiducial cosmology,
$D_* \approx 14$Gpc.  Since $j_\ell(x)$ peaks at $x\approx \ell +1/2$, the SW contributions
peak at $x=5/2$  or $k \sim 0.0002$ Mpc in Fig.~\ref{fig:quadadi}.  The temporal evolution for
the potential and quadrupole are shown for this mode
 in Fig.~\ref{fig:time}a.

The ISW effect arises 
from the change in the gravitational potential due to the 
dark energy.  In the adiabatic model, the smooth dark energy density accelerates the
expansion without enhancing the density perturbations and leads to a decay
of the potential at $z\simlt 1$  (see Fig.~\ref{fig:time}).   Naively, one would suppose
that a decay of $\Delta \Phi = -\zeta_i/10$ would be sufficient to cancel out the SW effect of
$\zeta_i/5$.  For superhorizon scaled adiabatic fluctuations in a flat universe the
conservation of the comoving curvature implies \cite{Bar80}
\begin{align}
\Delta \Phi \equiv& \Phi(k,a=1) - \Phi(k,a_{\rm m}) \nonumber\\
\approx & \left( { 2\over 5} - {H_0} \int_{a_{\rm m}}^1 {da ' \over H} \right) \zeta_i\,,
\label{eqn:phia}
\end{align}
where the approximation assumes $a_{\rm m} \ll 1$ is some epoch near the beginning of
matter domination so that contributions near the lower limit of the integral may be ignored.
Here $H=a^{-1}da/dt$ is the Hubble parameter.
In the fiducial model $\Delta \Phi \approx -0.14 \zeta_i$.  
However the distance to $z \simlt 1$ is much smaller than $D_*$ and that degrades
the efficiency with which the ISW effect contributes in Eqn.~(\ref{eqn:analytictransfer}).
The efficiency factor $j_2(kD)$ implies
that one requires factor of 10
greater change in the potential (or $\Delta \Phi \sim -\zeta_i$) to affect the quadrupole substantially (see Fig.~\ref{fig:time}).
 At the peak of the SW effect in $k$, the ISW effect has little effect
(see Fig.~\ref{fig:quadadi}).
 
 Contributions from the ISW effect in the quadrupole
 actually originate from scales where $k D_{\rm de} \approx 5/2$ where ``de" denotes
 the epoch of dark energy-matter equality.   Because the distance changes rapidly
 with redshift locally, the ISW effect is spread out across over a factor of ten in
 physical scale with a peak centered near $k=0.001$ Mpc$^{-1}$ (see Fig.~\ref{fig:quadadi}a).  
 At this wavenumber $j_2(kD)$ has a peak near $z=1$ corresponding to an efficient transfer of
 power.  Moreover since the decay occurs on the expansion time scale
 the oscillations in $k$ from $j_l(k D)$ in the ISW integrand of Eqn.~(\ref{eqn:analytictransfer})  are washed out in the transfer
 function. 
 Physically this reflects the cancellation of radial modes as photons travel in and out
 of decaying gravitational potentials along the line of sight.   
  
To lower the {\it predicted} value of the quadrupole, one can alter the fiducial model to lower 
the SW effect, the ISW effect or both.
Since both effects contribute nearly equally, reducing one or the other can at best halve the power.
 Of course, due to cosmic variance, it is possible that the {\it observed} quadrupole results from
 a lack of angular power in our given realization of the fiducial model.   
In Fig.~\ref{fig:cladi} we show the 68\% and 95\% cosmic variance confidence regions assuming
that $C_\ell^{\Theta\Theta}$ is distributed as a $\chi^2$ with $2\ell+1$ degrees of freedom around
the fiducial model.
However again, a simple lack of power on large
 physical scales for our last scattering (recombination) surface is not sufficient.  Unless 
 our local volume also lacks intermediate scale power as well, a chance occurrence of
 a low observed quadrupole would result from a chance cancellation of the SW and ISW effects.

The low multipole polarization and cross spectra
 provides key additional information to discriminate between alternatives.
 In the large scale
 limit, it is approximately
\begin{align}
T_\ell^E(k,\eta_0) & =  -{3\over 4} \sqrt{ (\ell+2)! \over (\ell-2)!} 
\int_{\eta_*}^{\eta_0} d\eta \dot \tau e^{-\tau} T_2^\Theta(k,\eta)
{j_\ell(k D) \over (kD)^2}\,,
\end{align}
where $\tau$ is the Thomson optical depth as measured from the observer. 
 Consider the transfer function of the polarization quadrupole.   Fig.~\ref{fig:quadadi}
shows that in the fiducial model it is generated before the temperature quadrupole
is modified by the ISW effect.
Consequently, the transfer function also peaks near the large scales of the
SW peak of $k \sim 0.0002$ Mpc$^{-1}$.  As Eqn.~(\ref{eqn:cl}) shows this overlap is also
the origin of the temperature-polarization cross correlation.
In Fig.~\ref{fig:pol}, we show the $EE$ and $\Theta E$ power spectra of the fiducial model.

If the explanation of the observed low quadrupole involves the dark energy, either
through a dynamical effect or chance cancellation, one would expect a polarization quadrupole and
hence $EE$ power
that is not anomalously low compared  with the fiducial model.   If on the other hand it involves an actual lack of
predicted long-wavelength power in the model or by chance, both spectra would be low.   Finally, if
the explanation involved only the reduction of the ISW effect and no modification of long wavelength
power, then the predicted $\Theta E$ cross power spectrum would also remain unchanged. 

\section{Dark Energy Models}
\label{sec:isocurvaturemodel}

The fact that in the fiducial adiabatic model the temperature quadrupole receives comparable
contributions from recombination and dark energy domination through the SW and ISW
effects raises the possibility that the low quadrupole originates in the dark energy sector.  In the
Appendix we present a detailed 
treatment of perturbation theory in general dark energy
models which provides the basis for results in this section. 

The essential element that defines the ISW effect in the fiducial model is that the dark energy
remains
smooth out to the horizon scale 
and hence does not contribute density fluctuations to the gravitational potential.  In general there
are two ways to alter this conclusion: modify the dynamics of the dark energy so that
dark matter fluctuations remain imprinted on the dark energy or modify the initial perturbations
in the dark energy sector.  Since the dark energy 
has made a negligible contribution to the net energy density
until recently, the latter represents contributions from an isocurvature initial condition.  
That dark energy isocurvature conditions can help to lower the quadrupole has recently 
been shown \cite{MorTak03}.  Here we present a general discussion on the requirements
of such a model.  

The first requirement is that an initial dark energy perturbation must survive evolution
in the radiation and matter dominated epoch and must 
remain correlated with the perturbations in the dark matter.  
The latter condition is required for the dark energy perturbations to cancel
the adiabatic ones.   

Let us take the dark energy to be a scalar field $Q$ with the canonical kinetic term and
a potential $V_Q$, i.e. quintessence.  We treat the more general case of k-essence
in the Appendix.    Given that a quintessence field has an effective sound speed $c_e=1$
[see Eqn.~(\ref{eqn:kessencesound})], coherence well inside the horizon and hence
cancellation with the adiabatic ISW effect is not possible.    The dark energy isocurvature
mechanism then must  operate on large scales to cancel the SW effect.

Recall that to achieve a coherent cancellation of the SW effect in the quadrupole one requires
either a change of $\Delta \Phi \sim -\zeta_i/10$ 
in the gravitational potential early on when $D \approx D_*$ or
a larger change $\Delta \Phi \sim -\zeta_i$ at  $z\simlt 1$
to compensate for the inefficiency of the transfer of power
to the quadrupole.   Given that observations require that $w_Q \sim -1$ today, the former
possibility is excluded unless $w_Q$ evolves substantially from its present value.

In a flat universe the comoving curvature evolves only in response to stress fluctuations in 
the combined or total (``T'') stress energy tensor of the components
(see Eqn.~\ref{eqn:bardeencurvature})
\begin{align}
\zeta(a,k) = &\zeta_i(k) - \int_{0}^a {d a' \over a'} {\delta p_T \over \rho_T + p_T } \nonumber\\
\approx & \zeta_i(k) -\int_{a_{\rm m}}^a {d a' \over a'} {\delta p_Q \over \rho_m }\,,
\label{eqn:bardeenevol}
\end{align}
where the approximation assumes $w_Q \approx -1$, $a\gg a_{\rm m}$, and the radiation 
and hence the anisotropic stress is negligible.
The generalization of Eqn.~(\ref{eqn:phia}) for the evolution in the Newtonian potential is [see e.g. \cite{HuEis99} Eqn.~(52)]
\begin{align}
\Phi(a,k) = & \zeta(a,k) - {H \over a} \int_{a_{\rm m}}^a {d a'  \over H} [\zeta -   {\delta p_T \over \rho_T + p_T }] \nonumber\\
                \approx &  \zeta(a,k) - {H \over a} \int_{a_{\rm m}}^a {d a'  \over H} [\zeta -   {\delta p_Q \over \rho_m }]\,.
\label{eqn:phievol}
\end{align}

Thus it requires a substantial pressure fluctuation 
to make an order unity change to gravitational potential during the dark energy dominated 
regime
\begin{equation}
{\delta p_Q \over \rho_m}(a=1,k) = {\Omega_Q \over \Omega_m} {\delta p_Q \over \rho_Q} (1,k)
= {\cal O}(\zeta_i) \,.
\end{equation}
Note that in the comoving gauge, adiabatic density fluctuations scale as $\delta \sim (k\eta)^2 \zeta_i$
[see Eqn.~(\ref{eqn:denvel})]
and are negligible outside the horizon.
Furthermore we shall see below that the order unity coefficient in front of $\zeta_i$ is in practice substantially
greater than unity.

This requirement severely limits the range of quintessence models which can affect the quadrupole.   
As shown in the Appendix, aside from transient initial
condition effects, an isocurvature perturbation to the quintessence field $\delta Q$ at best remains
constant outside the horizon and hence one requires a large initial fluctuation to the quintessence field.  

A constant superhorizon quintessence field fluctuation generically occurs if the background field itself is
nearly frozen by the Hubble drag so as to only experience a range in the potential
where
\begin{equation}
V_Q' \equiv {d V_Q \over d Q}
\end{equation}
can be approximated as constant.  More specifically, we require that the field not be in the
tracking regime or (\cite{SteWanZla99} see also Appendix)
\begin{equation}
\Gamma \equiv {V_Q'' \over V_Q} \left( {V_Q' \over V_Q } \right)^{-2} \simlt 1\,.
\label{eqn:trackercond}
\end{equation}
To see the consequences
for the energy density and pressure, note that aside from a transient decaying mode
the Klein-Gordon equation (\ref{eqn:KleinGordon}) has the solution
\begin{equation}
{d Q \over d\ln a} \approx - {2 \over 3(3+w_T)} {V_Q' \over H^2} \,,
\label{eqn:drag}
\end{equation}
where we have assumed an epoch during which $w_{T}=p_{T}/\rho_{T}$ the equation of state of the background is constant.
The dark energy density is the sum of the kinetic and potential components
\begin{equation}
\rho_Q = {1 \over 2} \left( H {d Q \over d\ln a} \right)^2 + V_Q \,.
\label{eqn:rhoq}
\end{equation}
Since $H^2 \propto \rho_T$ decreases with the expansion, if the field is potential energy
dominated today ($w_Q \sim -1$)
then it is potential energy dominated for the past expansion history.  The combination of
Eqn.~(\ref{eqn:drag}) and (\ref{eqn:rhoq}) shows that
this will be satisfied if the potential satisfies
\begin{align}
{1 \over 2} \left( V_Q'\over V_Q \right)^2 {V_Q \over H_0^2} & \simlt 1 \,,
\end{align}
or equivalently with 
\begin{equation}
\epsilon_Q = {1 \over 16\pi G} \left( {V_Q' \over V_Q}\right)^2 \,,
\label{eqn:epsilonq}
\end{equation}
and 
\begin{equation}
\Omega_{Q} \approx { 8 \pi G V_Q \over 3 H_0^2}  \,,
\end{equation}
the condition becomes
\begin{equation}
3 \Omega_Q \epsilon_Q \simlt 1\,.
\end{equation}
Moreover since the field only experiences a small range in the potential throughout the
whole expansion history, any underlying form of $V_Q$ that satisfies these requirements
will have the same phenomenology.  We find that in the context of the fiducial model
$\epsilon_Q \simlt 0.6$ is required for $w_Q(a=1) \simlt -2/3$.

Given potential energy domination, the energy and pressure fluctuations are related
to the field fluctuations as
\begin{equation}
\delta \rho_Q \equiv \delta_Q \rho_Q = -\delta p_Q = V_Q' \delta Q 
\end{equation}
and remain nearly constant during the evolution.  
The superhorizon 
evolution of the  comoving curvature  from Eqn.~(\ref{eqn:bardeenevol})  is given by 
\begin{equation}
\zeta(a,k) = \zeta_i(k) + {1 \over 3} {\delta_Q}{\Omega_Q \over \Omega_m} a^{3}
\end{equation}
and hence from Eqn.~(\ref{eqn:phievol}) 
\begin{align}
\Phi(a,k) =& \left[ 1 - {H \over a} \int_{a_{\rm m}}^a {da' \over H} \right] \zeta_i \nonumber\\
   &+ \left[ 1 - 4{H\over a}\int_{a_{\rm m}}^a {da' \over H} {a'}^3 \right] {1 \over 3} \delta_Q {\Omega_Q \over \Omega_m}
\label{eqn:phisoln}
\end{align}
where the two pieces are the adiabatic term and the isocurvature term.  
The integrals in Eqn.~(\ref{eqn:phisoln}) can be expressed in terms of hypergeometric functions.
In the fiducial model 
\begin{equation}
\Phi(a=1,k) = 0.46 \zeta_i + 0.12 \delta_Q\,.
\end{equation}
Combined with the requirement that $\Delta \Phi \approx -\zeta_{i}$ this relation 
implies that we should set the initial dark energy
fluctuations to be $\delta_{Qi} \approx -8 \zeta_i$ to cancel the quadrupole in the fiducial 
model.  

\begin{figure}[tb]
\centerline{\epsfxsize=3.2in\epsffile{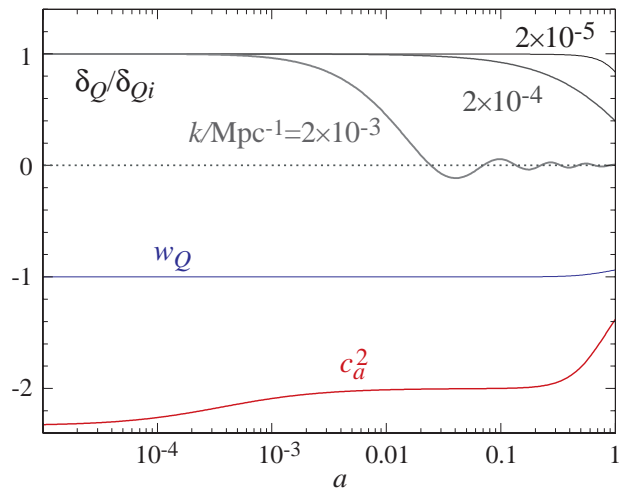}}
\caption{\footnotesize Isocurvature dark energy density perturbation
evolution for a nearly frozen background quintessence field (see
text). On superhorizon scales the comoving gauge density perturbation
remains frozen at nearly its initial value.  On subhorizon scales, the
pressure gradients associated with field fluctuations cause the
density perturbation to oscillate and decay.  Also shown are the dark
energy equation of state $w_Q = p_Q/\rho_Q$ and the adiabatic sound
speed squared $c_a^2 = \dot p_Q /\dot \rho_Q$ (not to be confused with the effective
sound speed $c_e=1$) which indicate the field is
Hubble drag dominated until recently (see Appendix).}
\label{fig:qevol}
\end{figure}

While this condition
is roughly correct, 
the scales that are responsible for the quadrupole in the SW effect ($k\sim 0.0002$ Mpc$^{-1}$)
are on the horizon scale today.  Since the quintessence field has a sound horizon equal to the
horizon, the fluctuations in these modes will have already begun to decay from their 
initial values.  In Fig.~\ref{fig:qevol} we show the time evolution of the dark energy
density perturbation.  Because the argument  above might appear to require  $w_Q=-1$ exactly, 
we have chosen to illustrate the behavior in a model with $w_{Q}(a=1)=-0.94$ 
and $\epsilon_{Q}=0.18$ with the other parameters equal to their fiducial values.  Since
$w_Q \rightarrow -1$ rapidly with redshift, the change in $D_*$ from the fiducial model is
only $0.4\%$.  As the background evolution is nearly indistinguishable from the
true fiducial model, we will employ this choice for the isocurvature analog of the
fiducial model.

In practice we have taken a potential $V_Q =m_Q^2 Q^2/2$ with $m_Q
= 10^{-42}$ GeV and an initial position consistent with $\epsilon_Q$ and $dQ/d\ln a =0$ initially.  
For scales near the peak of the SW effect in the quadrupole
the density perturbation has decayed by about a factor of two by $z\sim1$.
Consequently an initial density perturbation of 
\begin{equation}
S_{i} \equiv {\delta_{Qi} \over \zeta_{i}} = -15
\label{eqn:cancelcondition}
\end{equation}
should be optimal for reducing the SW quadrupole.  
We call models with this type of fully correlated adiabatic and isocurvature models
``A\&I'' models.

In Fig.~\ref{fig:time} we show the effect of this initial condition on the gravitational potential.  In this case the change in the gravitational potential $\Delta \Phi 
\approx -\zeta_i$ and one achieves the desired effect of eliminating the SW effect in the quadrupole.

\begin{figure}[tb]
\centerline{\epsfxsize=3.4in\epsffile{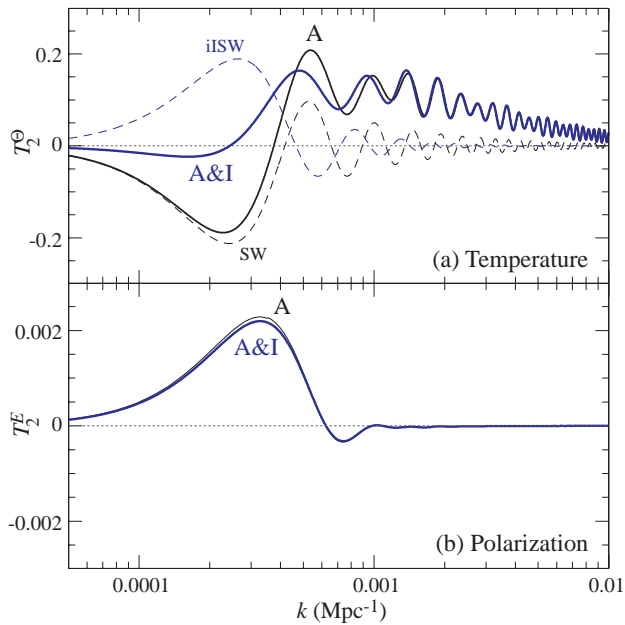}}
\caption{\footnotesize
Quadrupole transfer functions in the fiducial adiabatic model (``A") and a model with additional
isocurvature perturbations of $S_{i}=-15$ (``A\&I").
The isocurvature ISW effect (iISW) cancel the SW effect for the
temperature quadrupole from large scales while leaving the polarization nearly unchanged. }
\label{fig:quadiso}
\end{figure}

The reduction of the SW quadrupole must not come at the expense of an enhancement in
the ISW contributions to the quadrupole.  Fortunately, this is a natural consequence of
the effective sound speed of the scalar field $c_e=1$.
Scales near the peak contribution of the adiabatic ISW effect are well within the horizon
by dark energy domination.   Consequently as can be seen in Fig.~\ref{fig:qevol} any initial
isocurvature perturbation in the dark energy will have decayed before dark energy domination.
In Fig.~\ref{fig:time}, we show the evolution of the potential given the initial conditions of
Eqn.~(\ref{eqn:cancelcondition}).  Note that despite the large initial isocurvature perturbation
it has almost no effect on the potential evolution and hence the contributions to the quadrupole.

\begin{figure}[tb]
\centerline{\epsfxsize=3.4in\epsffile{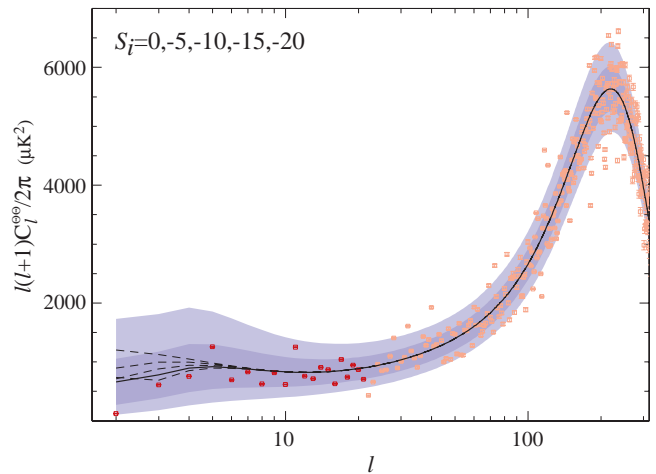}}
\caption{\footnotesize
Temperature power spectrum of the A\&I model with several choices for the initial isocurvature
amplitude compared with the WMAP data.  The shaded region corresponds to the 68\% and
95\% cosmic variance confidence region of the fiducial A\&I model with $S_i=-15$ (solid line).}
\label{fig:cl}
\end{figure}

In Fig.~\ref{fig:quadiso}, we show the temperature and polarization transfer functions of the
isocurvature model.  The isocurvature ISW effect almost perfectly cancels the
SW effect on large scales leaving a quadrupole that consists almost solely of the
adiabatic ISW effect.  Note that the isocurvature conditions leave the polarization essentially
unmodified as expected since they only change the potentials at  $z \simlt 2$ 
in Fig.~\ref{fig:time}.  

\begin{figure}[tb]
\centerline{\epsfxsize=3.3in\epsffile{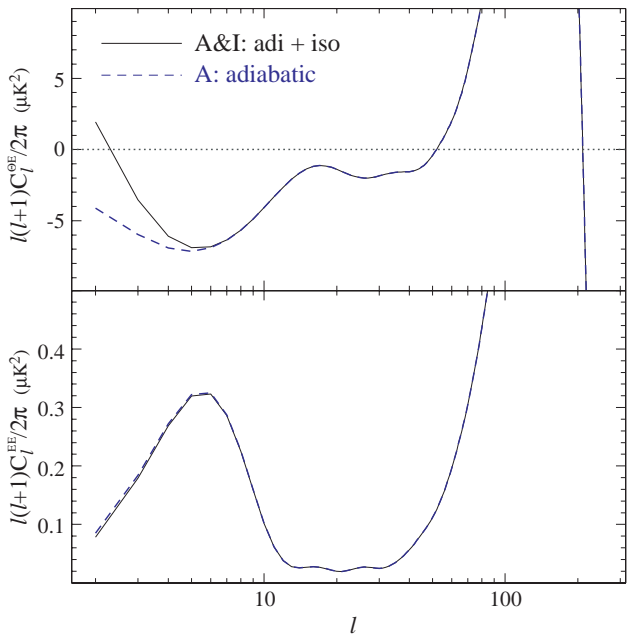}}
\caption{\footnotesize
The polarization $EE$ and cross correlation $\Theta E$ spectra for the fiducial adiabatic
model  compared with the fiducial A\&I model of Fig.~\ref{fig:cl} with $S_{i}=-15$.  Although the
$EE$ spectrum remains essentially unchanged, the correlation is modified 
at the 
low multipoles
due to the reduction of the 
large scale contributions to the quadrupole from the SW effect which are the source of
the correlation in the adiabatic model.}
\label{fig:pol}
\end{figure}

In Fig.~\ref{fig:cl} we show the temperature power spectrum of this model for several
choices of $S_{i}$ with all other parameters held fixed.  Note that for
$S_{i} \approx -15$ the suppression is fairly sharp around the quadrupole in
spite of scale invariance in both the adiabatic and isocurvature initial conditions.
  The polarization auto ($EE$) and cross  ($\Theta E$) spectra are shown in Fig.~\ref{fig:pol}.  
Notice that although the $EE$ spectrum remains nearly unchanged, the $\Theta E$ spectrum
has a reduced cross correlation with a sign change at the quadrupole. (Note that our sign
convention for $E$ is opposite to CMBFAST.)    
Because the cross spectrum in the adiabatic case arises
from the correlation between the SW quadrupole and the polarization, it is affected by the
cancellation of the SW quadrupole as the product of the two transfer functions in
Fig.~\ref{fig:quadiso} show.

\begin{figure}[tb]
\centerline{\epsfxsize=3.4in\epsffile{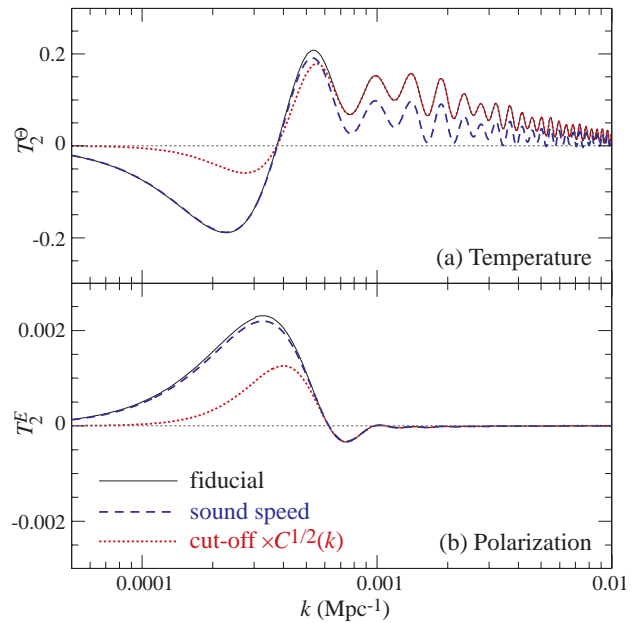}}
\caption{\footnotesize
Alternate models for the low quadrupole.   The adiabatic ISW effect may be reduced
by lowering the effective sound speed of the dark energy,
here to $c_e = 0.03$.  The SW effect can be largely eliminated 
through a cut off factor $C(k)$ in the initial adiabatic power spectrum, 
here chosen to remove power for  $k< k_{\rm cut} =0.0005$ 
Mpc$^{-1}$.  Although the latter
does not modify the transfer function, we have illustrated its effects by
showing $T_\ell C^{1/2}(k)$.  Note that the cut off affects the polarization whereas the
sound speed does not.}
\label{fig:quadalt}
\end{figure}

\begin{figure}[tb]
\centerline{\epsfxsize=3.4in\epsffile{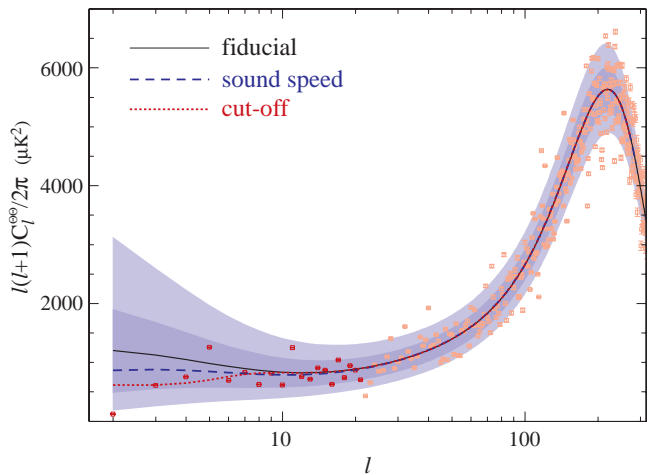}}
\caption{\footnotesize
Temperature power spectra of the sound speed and cut off models of Fig.~\ref{fig:quadalt}.  The
sound speed model lowers the adiabatic ISW contributions to the quadrupole.  The cut off
model reduces the SW contributions to the quadrupole.}
\label{fig:clalt}
\end{figure}

\begin{figure}[tb]
\centerline{\epsfxsize=3.3in\epsffile{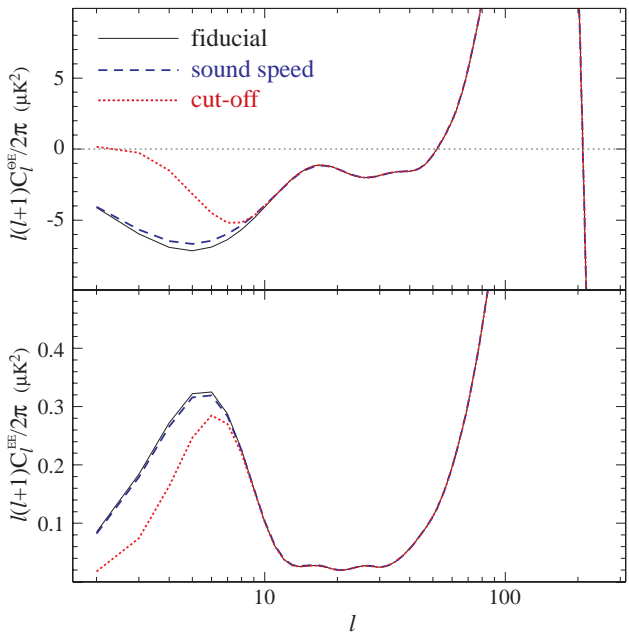}}
\caption{\footnotesize
Polarization spectra for the sound speed and cut off models of Fig.~\ref{fig:quadalt}.  The sound
speed model changes both spectra negligibly whereas the cut off model suppresses power
in both spectra simultaneously.}
\label{fig:polalt}
\end{figure}

It is interesting to compare the isocurvature method of lowering the quadrupole to other
possible solutions.  A related mechanism involves modifying the sound speed of the
dark energy \cite{Hu98,DedCalSte03,Erietal02,BeaDor04}.    Here one retains adiabatic initial
conditions but modifies the effective sound speed of the dark energy 
(see Appendix).  The dark energy then contributes density fluctuations
between the horizon and the sound horizon and reduces the adiabatic ISW effect by suppressing
the decay of the gravitational potential.  
In Fig.~\ref{fig:quadalt}, we show the transfer functions for a model with $w_Q=-2/3$,
$c_e = 0.03$ with $\Omega_Q=0.37$ and $\Omega_m h^2$ and $\Omega_b h^2$ held
fixed to the fiducial values such that $D_*$ and the shape of the peaks remain the same
as in the fiducial model.  Note that unlike in the isocurvature case, the SW contributions to the
quadrupole are unmodified but the adiabatic ISW contributions are reduced.  The resulting
effect on the temperature power spectrum is shown in Fig.~\ref{fig:clalt}.   However,
like the isocurvature case, the polarization is largely unchanged since
the dark energy again only affects low redshifts.    A qualitative difference appears
in the cross power spectra, which remains unchanged in this case (see Fig.~\ref{fig:polalt}).

Finally, the quadrupole can be lowered by removing power on scales associated
with the SW effect.   Here we take the model \cite{ConPelKofLin03} 
\begin{align}
\Delta^2_{\zeta_i}(k) =& \Delta^2_{\zeta_i}(k) |_{\rm fid} C(k)\,, \nonumber\\
C(k) = &1 - e^{-(k/k_{\rm cut})^{n_{\rm cut}}}\,,
\end{align}
with $k_{\rm cut} = 0.0005$ Mpc$^{-1}$ and $n_{\rm cut}=3.35$.  The transfer functions
for this model are the same as the fiducial model but for illustrative effect we plot them
as $T_\ell(k) C^{1/2}(k)$ in Fig.~\ref{fig:quadalt}.  Like the isocurvature model, the quadrupole
in Fig.~\ref{fig:clalt} is suppressed due to the elimination of the SW quadrupole.  Unlike
the isocurvature case, both the $EE$ and $\Theta E$ spectra are suppressed since
the temperature quadrupoles are also absent during reionization (see Fig.~\ref{fig:polalt}).

\section{Likelihood Analysis}
\label{sec:likelihood}
In this section we check whether the adiabatic plus isocurvature (A\&I) model
of the previous section is favored by the WMAP data
\cite{Ben03}. For simplicity we will assume a scale invariant
spectrum, zero background curvature
and zero neutrino mass. Then the maximum likelihood fit with no dark
energy perturbations is given in Table~\ref{maxlik}.
\begin{table}
\caption{\footnotesize \label{maxlik} 
Scale Invariant $\Lambda$CDM Model Parameters: WMAP $\Theta\Theta$ $\&$ $\Theta E$ Data. 
}
\begin{tabular}{lc}
\hline \hline Parameter & Maximum Likelihood\\
\hline
Matter / Critical Density, $\Omega_m$ & 0.27\\
Baryon / Critical Density, $\Omega_b$ & 0.046\\
Hubble Constant, $h$& 0.72\\
Optical Depth, $\tau$ & 0.17\\
Initial Amplitude, $\delta_{\zeta_{i}}$&5.07$\times$ $10^{-5}$\\
\hline
\end{tabular}
\end{table}
When the spectral index is fixed the only parameter with enough
freedom to significantly effect the low multipoles is the optical
depth, $\tau$. 

Keeping the other parameters in Table~\ref{maxlik}
fixed we vary the magnitude $S_{i}$ of the primordial isocurvature
perturbation in dark energy relative to the comoving curvature 
between $-60 < S_{i} < 25$ and the optical depth $0<\tau<0.29$ on a
$28\times 30$ grid. The amplitude in Table~\ref{maxlik} is scaled
by $\exp(\tau-0.17)$; recall that the power spectrum is scaled as the square of this
quantity [see Eqn.~(\ref{eqn:fiducialspect})]. This is to take into account the well known
degeneracy between $\tau$ and the amplitude.
For each grid point the likelihood was evaluated
using the software provided by WMAP \cite{Hin03,Ver03} and the resulting
two dimensional surface was interpolated. A uniform prior
in both parameters was assumed.

\begin{figure}
\centerline{\epsfxsize=3.0in\epsffile{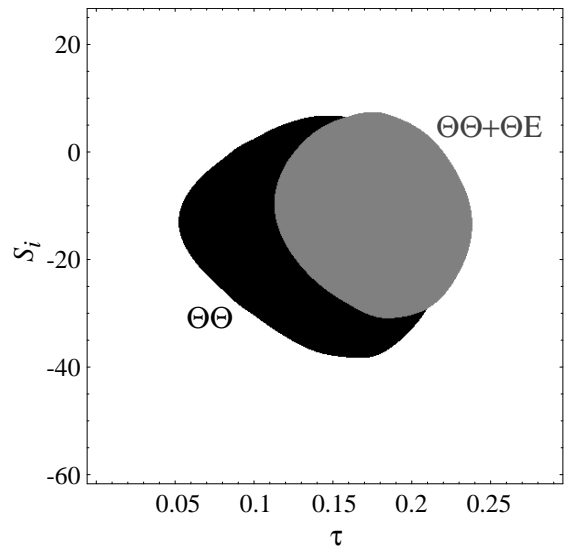}}
\caption{\footnotesize
\label{pdf2d} 
Areas enclosing 95\% of the probability for the dark energy
isocurvature perturbation ($S_{i}$) and the optical
depth ($\tau$). The black area is using only the $\Theta\Theta$ WMAP data and the
gray area uses both the $\Theta\Theta$ and $\Theta E$ WMAP data.}
\end{figure}

The maximum likelihood points are $S_{i}=-15.3$
and $\tau=0.16$ for WMAP $\Theta\Theta$ data only and  $S_{i}=-12.1$
and $\tau=0.18$ for WMAP $\Theta\Theta$ and $\Theta E$ data.
The $\chi^2$ for these parameter values and other models are giving
in Table~\ref{chi2}.
The reduction in $\chi^2$ is achieved by only modifying the 
first few multipoles of the predicted spectrum. 
\begin{table}
\caption{\footnotesize \label{chi2}
Best fit (assuming scale invariance) $\chi^2$ and degrees of freedom (DOF) values using WMAP data.
}
\begin{tabular}{llcc}
\hline \hline Model & Data & $\chi^2$ & DOF\\
\hline
Fiducial (Table~\ref{maxlik})& $\Theta\Theta$&         976	     & 894 \\
$c_e = 0.03$ & $\Theta\Theta$&         973          & 893    \\
Cut-off ($k_{\rm cut}=0.0005$Mpc$^{-1}$)   & $\Theta\Theta$&    972      & 893 \\
Adiabatic \& Isocurvature    & $\Theta\Theta$&         972          &	 893  \\
\hline
Fiducial (Table~\ref{maxlik})& $\Theta\Theta+\Theta E$&         1430	     & 1343 \\
$c_e = 0.03$ & $\Theta\Theta+\Theta E$&   1428              & 1342	    \\
Cut-off ($k_{\rm cut}=0.0003$Mpc$^{-1}$) &$\Theta\Theta+\Theta E$& 1428     &	1342 \\
Adiabatic \& Isocurvature  & $\Theta\Theta+\Theta E$&         1427          & 1342	    \\
\hline
\end{tabular}
\end{table}
 Areas enclosing 95\% of the
probability with and without the $\Theta E$ data are shown in
Fig.~\ref{pdf2d}. The area perimeters lie on constant probability
contours. 
As can be seen from Fig.~\ref{pdf2d}, the weight of the probability distribution favors $S_{i}<0$, i.e.
an isocurvature density perturbation that is negatively correlated with the comoving curvature
perturbation, $\zeta_{i}$. This preference comes from the fact that the lowered temperature quadrupole in
such a model better fits the observations.  
The addition of the $\Theta E$ data reduces the favored magnitude of
the isocurvature perturbation $S_{i}$ and slightly increases that of $\tau$
as $\Theta E$ is suppressed in the model but not
in the data \cite{DorHolLoe03}.  Note that the restriction to scale invariant models
does not affect the relative improvement that $S_i$ makes for a given $\tau$ since
across the scales relevant to the first few multipoles any nearly scale invariant
model will have the same shape.  Allowing the tilt to vary does allow low $\tau$
models if only the $\Theta\Theta$ data is employed but these are eliminated
with the inclusion of $\Theta E$ data (c.f. \cite{Spe03} and \cite{Hanetal04}). 

\begin{figure}
\centerline{\epsfxsize=3.2in\epsffile{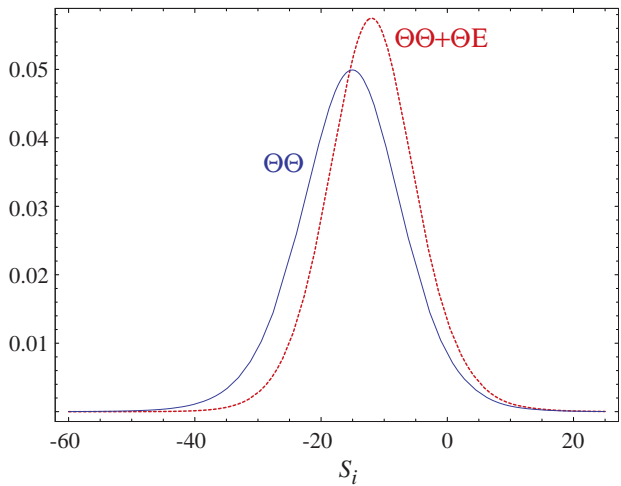}}
\caption{\footnotesize
\label{pdf} 
Probability density functions for the dark energy isocurvature
perturbation using WMAP $\Theta\Theta$ (solid) and WMAP $\Theta\Theta$ and $\Theta E$ data (dashed).
}
\end{figure}

One dimensional probability distributions obtained by integrating out
$\tau$ are displayed in Fig.~\ref{pdf}.
The mean and 68\% confidence regions for $S_{i}$ are
$-15.2 \pm 8.2$ for WMAP $\Theta\Theta$ data only and $-11.8 \pm
7.1$ for WMAP $\Theta\Theta$ and $\Theta E$ data. The probability of $S_{i}>0$ is 0.04 using WMAP $\Theta\Theta$ data only and 0.06 using WMAP $\Theta\Theta$ and
$\Theta E$ data. This is in sharp contrast to the estimates of the amount of
totally correlated CDM, baryon or neutrino isocurvature perturbation as estimated
in \cite{GorLew03,GorMal04} where the probability for a positive isocurvature
perturbation was roughly 50\%, i.e.\ the isocurvature perturbation was
not significantly different from zero. The reason for this is that
correlated non-dark energy isocurvature  perturbations have a similar
effect on a much larger range of low $\ell$ multipoles as they rely on a
simple reduction of the SW effect. The dark energy introduces
a scale associated with the current horizon and so the dark energy isocurvature ISW effect
can only reduce the SW effect on the large scales associated with the
quadrupole.

\section{Inflationary Context}
\label{sec:inflation}
 
As we have shown, it is essential that the dark energy isocurvature
perturbation is anti-correlated with the adiabatic perturbation so that its
ISW effect can coherently cancel the SW effect. Correlated or
anti-correlated adiabatic
and isocurvature perturbations result when the adiabatic curvature perturbation itself 
is generated by the isocurvature perturbation
\cite{Lan99,GorWanBasMaa01}.  Ref.~\cite{MorTak03} suggested a three field model based on
the curvaton mechanism \cite{EnqSlo02,LytWan01,MorTak01} in order to
 produce  anti-correlated isocurvature dark energy
perturbations. 

Another way of establishing the anticorrelation  is through the variable
decay mechanism \cite{DvaGruZal03,Kof03} where there is a second light
scalar field during inflation which determines the decay rate
$\Gamma$ of the inflaton. In this method, the resulting comoving
curvature perturbation is given by 
\begin{equation}
\zeta_{i} =-{ 1\over 6}{\delta
\Gamma \over\Gamma} \,.
\end{equation}
If we associate the variable decay field with the
quintessence field, $Q$, and take $\Gamma \propto Q$ then
$\zeta_{i} =-(1/6)\delta Q/Q$. Then, if we further take the quintessence
potential to be $V_Q \propto Q^2$, we have $\delta_{Q_{i}}=2\delta
Q/Q.$ Therefore 
\begin{equation}
\delta_{Q_{i}} =-12\zeta_{i}\,,
\label{eqn:modelpredict}
\end{equation} 
which was the mean
value found in \S\ref{sec:likelihood} when both WMAP $\Theta\Theta$
and $\Theta E$ data were used.

Unfortunately, a problem with this and other inflationary scenarios
arises because of the gravitational waves generated during inflation. 
Both the scalar field and gravitational waves are nearly massless degrees
of freedom during inflation and hence acquire closely related quantum fluctuations.
The contribution of the
gravitational waves to the $\Theta\Theta$ spectrum for $\ell \gg 1$ is \cite{Sta85}
\begin{equation}
 {\ell(\ell+1) \over 2\pi }C_\ell^{\Theta\Theta} \Big|_{\rm grav} \approx 2\pi  G \left(
\frac{\Hinf}{2\pi}\right)^2\,,
\end{equation}
 where $\Hinf$ is the Hubble parameter
during inflation.  On the other hand, Eqn.~(\ref{eqn:analytictransfer})
implies that the curvature fluctuations induce a SW effect of
\begin{equation}
{\ell(\ell+1) \over 2\pi}C_\ell^{\Theta\Theta}  
\Big|_{\rm SW}  \approx {1 \over 25} \Delta_{\zeta_{i}}^{2}\,.
\end{equation}
Now $\Delta_{\zeta_{i}}^{2}$ is related to the inflationary power spectrum of $\delta Q$,
\begin{equation}
\Delta_{\delta Q}^{2} = \left( {\Hinf \over 2\pi } \right)^{2}
\label{eqn:fieldpower}
\end{equation}
through
\begin{eqnarray}
-12 \zeta_{i} &\approx&{\delta_{Qi}} \approx
 {V_Q' \over V_Q } \delta Q \nonumber \\
&=& \sqrt{16\pi G \epsilon_Q} \delta Q\,,
\end{eqnarray}
where $\epsilon_Q$ was defined in Eqn.~(\ref{eqn:epsilonq}). Unlike inflaton curvature fluctuations, these curvature fluctuations are suppressed with a small $\epsilon_Q$ because the cancellation
condition involves the energy density fluctuation in $Q$.
Therefore
\begin{equation}
\frac{\ell(\ell+1)}{2\pi}C_\ell \Big|_{\rm SW}  \approx \frac{G\pi}{225}\epsilon_Q \left(
\frac{\Hinf}{2\pi}\right)^2.
\end{equation}
The predicted ratio of
gravitational wave to SW contributions is $r\approx
450/\epsilon_Q$.

As discussed in \S \ref{sec:isocurvaturemodel}, 
an equation of state today of
$w_Q \approx -1$ requires a small slow roll parameter for the
quintessence ($\epsilon_Q \simlt 0.6$) at the field position today.  
The current observational limit is about $r<1$
\cite{Pei03}. Therefore, the isocurvature perturbations in a
 nearly frozen field model for the
quintessence field could not have originated from this inflationary mechanism.

Note that the problem of excess gravitational waves is independent of
the precise potential or inflaton decay rate relation as it arises from
the phenomenologically required condition Eqn.~(\ref{eqn:modelpredict}). It applies
to any model (including the curvaton model proposed in
Ref.~\cite{MorTak03}) for which the field fluctuations $\delta Q$ at
dark energy domination are directly associated with the inflationary
fluctuations of a light canonical scalar field [Eqn.~(\ref{eqn:fieldpower})].

The only way to evade this conclusion is for $\delta Q$ to grow by a factor of
30 or more between horizon crossing during inflation and today. It is
not clear how to do this for a canonical scalar field whose perturbations
were generated during inflation.

\section{Discussion}

We have explicated the mechanism by which scale invariant isocurvature perturbations
in the dark energy can lead to a sharp suppression of the quadrupole temperature anisotropy \cite{MorTak03}.
The Integrated Sachs-Wolfe (ISW) effect in the dark energy dominated epoch 
can coherently cancel the Sachs-Wolfe (SW) effect from recombination if its energy density
fluctuations are set to be strongly anticorrelated with the initial curvature fluctuations.  We call these A\&I models. 
Assuming a scale invariant spectrum and flat background, we
detect the presence of A\&I perturbations
at the 95\% confidence level.

As is
well known \cite{AbrFin01,BarCorLidMal04,FerJoy97} (see also Appendix),
isocurvature modes in
the dark energy rapidly decay for tracking models.
Models in which the scalar field
and its isocurvature perturbations are essentially frozen
\cite{BarCorLidMal04} due to a shallow potential slope relative to
the Hubble parameter are better suited.  We have shown that
the requirements on such a model comes mainly through the
 quintessence ``slow roll'' parameter $\epsilon_{Q}$.  
 
 As for the origin of the A\&I correlation, 
 we can associate the dark energy with a
variable decay rate of the inflaton. We find that the right level of
isocurvature perturbations is naturally predicted.  Unfortunately, the
requirement of a shallow slope of the potential or small $\epsilon_{Q}$ implies
too high  a level of gravitational waves.  For this model to work
some mechanism is required for amplifying the field fluctuations by a factor of 30 or
more between inflation and dark energy domination. 

A\&I models should be contrasted with those that introduce a cut off scale
to the perturbations. A conceptual problem of the latter class
 is that it introduces the sharp reduction at the quadrupole ``by hand''.  In other
 words, there is a new coincidence
problem between the cut-off or topology scale and the horizon today.  There is also
no significant evidence for features at any other scale (e.g. \cite{MukWan04,Han04}).
Phenomenologically, by eliminating the
perturbations altogether, the cut off models also eliminate the source of
large angle polarization.  Thus, the cut-off and A\&I models are potentially
distinguishable from the polarization autocorrelation (or $EE$) spectrum.  On the other
hand both
models predict a reduced temperature polarization cross correlation.   

In principle, 
these alternatives can also be distinguished by large scale measurements of
the density field as a function of redshift and its
ISW correlation with the CMB 
\cite{KesKamCor03,HanMer04,Afs04,HuScr04}, e.g.
through high redshift galaxies \cite{CriTur96,BouCri04,FosGaz04,AfsLohStr04} or
cosmic shear \cite{Hu01c,Hut01,Son04}, but measuring the small signals involved will require
exquisite control over systematics in the surveys.

Both cut off and A\&I models operate by reducing the SW contributions
to the quadrupole.  Unfortunately, the SW effect only contributes approximately
half of the quadrupole in adiabatic models with a cosmological constant.  The
remaining portion comes from the ISW effect and receives contributions
across  a wide range of subhorizon scales.  
Neither A\&I nor cut off models can suppress the ISW effect at the quadrupole.
In the former, the high sound speed of the dark energy prevents substantial
density perturbations on subhorizon scales.  In the latter, a cut-off  on small enough
scales to affect the ISW effect would remove too much small scale power and
distort the higher $\ell$ multipoles.    

The adiabatic ISW effect can be modified by changing the
 dynamics of the dark energy by lowering its effective sound speed.  Alone
 it only amounts
in a fairly small reduction if the other multipoles are not to be
adversely affected.  This model is also distinguished from the A\&I and cut off models
in that it affects neither the auto nor the cross correlation spectra of the polarization.
This is because it is the SW quadrupole that is responsible for the polarization 
and hence its correlation with the temperature. 

In combination with an anticorrelated A\&I model, a sound speed modification alters the
multipole at which the SW effect is canceled.  We find that the canonical 
sound speed $c_e=1$ is nearly optimal in producing a sharp suppression at the 
quadrupole though a slight increase in the sound speed can actually
lead to a slightly sharper reduction.

Since the first COBE detection,
the low quadrupole temperature anisotropy in the CMB has provided a tantalizing
hint that new physics may be hovering on the horizon scale.  
With upcoming polarization auto and cross correlation data 
from WMAP, we may soon more than double the information on this intriguing problem.
All of the alternatives discussed here have distinct, albeit cosmic variance limited,
predictions for these spectra.  The predictions
 are especially distinct at the quadrupole and octopole but it remains to be seen
 how well these large-angle polarization fluctuations can be separated from galactic
foregrounds and instrumental effects.

\smallskip{\it Acknowledgments:}  We thank D. Eisenstein and L. Kofman for useful discussions.
CG was supported 
by the KICP under NSF PHY-0114422; WH by the DOE and the Packard Foundation. 
\appendix
\section{Dark Energy Perturbations}

\newcommand{\whit}[1]{#1}
\newcommand{\yell}[1]{#1}
\newcommand{\mage}[1]{#1}
\newcommand{\oran}[1]{#1}
\newcommand{\eqnsize}{}
\newcommand{\mytitle}[1]{{\bf #1}} 
\newcommand{\mybullet}[1]{#1}
\newcommand{\mynbullet}[1]{#1}
\newcommand{\vertsp}{}
\newcommand{\potential}{{A}}
\newcommand{\shift}{{B}}
\newcommand{\curvature}{{H}_L}
\newcommand{\shear}{{H}_T}

\subsection{Covariant Conservation}

Following  \cite{Bar80,KodSas84}, 
let us define the most general perturbation to the Friedmann Robertson Walker (FRW)
metric as
\begin{align}
\yell{g^{00}} &= -a^{-2}(1-2 \yell{\potential})\,, \nonumber\\
\yell{g^{0i}} &= -a^{-2} \yell{\shift^i}\,,  \nonumber\\
\yell{g^{ij}} &= a^{-2} (\gamma^{ij} -
        2 \yell{\curvature} \gamma^{ij} - 2 \yell{\shear^{ij}})\,,
\label{eqn:metric}
\end{align}
where $\gamma_{ij}$ is the FRW 3-metric of constant (comoving) curvature 
$K=H_0^2 (\Omega_{T}-1)$.  Likewise let us define the most general stress-energy
parameterization
of an energy density component $Q$
\begin{align}
\yell{T^0_{\hphantom{0}0}} &= -\rho_Q, \nonumber\vertsp\\
\yell{T^0_{\hphantom{0}i}} &= q_{Q\, i} \,, \nonumber\vertsp\\
\yell{T^i_{\hphantom{i}j}} &= p_Q  \delta^i_{\hphantom{i}j}
        + \yell{\Pi_{Q}} {}^{i}_{\hphantom{i}j}\,, \vertsp
\label{eqn:stressenergy}
\end{align}
where $\rho_Q$ is the energy density, $q_{Q\, i}$ is the momentum density,
$p_Q$ is the pressure or isotropic stress and $\Pi_{Q}{}^{i}_{\hphantom{i}j}$is the anisotropic stress of $Q$.  The anisotropic stress is defined as the trace free portion of the
stress such that 
$\Pi_{Q}{}^{i}_{\hphantom{i}i}$=0.  If $Q$ is to 
dominate the expansion at any time then $q_{Q}$ and $\Pi_{Q}$ must vanish in
the background due to isotropy.

If $Q$ is non-interacting or ``dark" then this stress energy tensor is
covariantly conserved.  Conservation then leads to general equations of motion for
the stress-energy components.  
Retaining terms linear in the metric fluctuations where they combine with
stress energy components $(\rho_{Q},p_{Q})$ in the background, we can reduce
$\nabla^\nu T^\mu_{\hphantom{0}\nu}=0$
to 
\begin{align}
\left[{d \over d\eta} + 3 {\dot a \over a}\right] \yell{\rho_Q} 
=&  - 3{\dot a \over a} \yell{p_Q}-  \nabla_{i} q^{i}_{Q} \nonumber\\
& -
(\rho_Q+ p_Q)( \nabla_{i}B^{i} +  3 \yell{\dot H_L})\,, \nonumber\\
\left[ {d \over d\eta} + 4{\dot a \over a}\right]  q_{Q\, i}
=& -\nabla_{i} p_{Q} - \nabla_{j} \yell{\Pi_Q} {}^j_{\hphantom{j}i}
 \nonumber\\
& - (\rho_Q+ p_Q) \nabla_i \yell{A} \,.
  \label{eqn:conservation}
\end{align}
The conservation equations represent a general but incomplete description
of the dark energy as they leave the spatial stresses
$p_{Q}$ and $\Pi_{Q}$ unspecified.  

\subsection{Equation of State}

The isotropic stresses may be 
rewritten in terms of an equation of state 
\begin{align}
p_{Q}(\eta,x_{i}) = w_{Q}(\eta,x_{i}) \rho_{Q}(\eta,x_{i})\,,
\end{align}
where $w_Q$ is in general a function of position and time.

For the scalar degrees of freedom it is useful to recast Eqn.~(\ref{eqn:conservation}) in 
terms of an expansion of the fluctuations into scalar harmonic modes defined by the  
complete set of
eigenfuctions $Y$ of the Laplace operator
\begin{equation}
\nabla^{2} Y = -k^{2}Y \,.
\end{equation}
The spatial fluctuations in each mode are given by four mode amplitudes
${\delta_Q, \delta w_Q, 
u_Q, \pi_Q}$ as
\begin{align}
 {\delta \rho_{Q}} \equiv& \delta_{Q} Y \rho_Q\,, \nonumber\\
 {\delta p_{Q} } \equiv&  (w_{Q} \delta_{Q} + \delta w_{Q})Y \rho_Q\,, \nonumber\\
{q_{Q\, i} } \equiv & u_Q ( -k^{-1} \nabla_{i} Y) \rho_Q\,, \nonumber\\
 {\yell{\Pi_{Q}} {}^{i}_{\hphantom{i}j} }
 \equiv & \pi_Q ( k^{-2} \nabla^{i}\nabla_{j} Y+{1\over 3}  \delta^i_{\hphantom{i}j}Y)  p_Q\,.
 \label{eqn:fluidfluct}
\end{align}
Note that for a spatially flat metric $Y = e^{i {\bf k} \cdot {\bf x}}$ and the mode
amplitudes are the Fourier coefficients of the fields.
In the literature (e.g. \cite{Bar80}), one often defines instead a quantity 
$V_Q$, related to the bulk velocity or energy flux, such that $u_Q = (1+w_Q)(V_Q-B)$. 
In Eqn.~(\ref{eqn:fluidfluct}) and throughout the remainder of this
 section $\rho_Q$, $p_Q$, and $w_Q=p_Q/\rho_Q$ are to be reinterpreted as 
 the average or background quantities and depend only on time. Hereafter we will typically
 omit the spatial harmonics $Y$ by assuming a harmonic
 space representation of the perturbations.
 
The conservation equations then become
\begin{align}
 \dot \delta_Q =& -3 {\dot a \over a} \delta w_Q - k u_Q   - (1+w_Q)(k B + 3 \dot H_L) \,,\nonumber\\
 \dot u_Q = & {\dot a \over a} (3w_Q-1)u_Q  + k (w_Q \delta_Q + \delta w_Q) 
  \nonumber\\
 &- {2 \over 3} w_Q  (1-3K/k^2) k \pi_Q + (1+w_Q) k A \,,
 \label{eqn:perturbedgeneral}
 \end{align}
where we have 
used the conservation of energy relation in
the background [see Eqn.~(\ref{eqn:conservation})]
\begin{align}
\dot \rho_Q = -3 {\dot a \over a} (1+w_Q) \rho_Q \,.
\label{eqn:backgroundrho}
\end{align}
Here we have followed the same convention for the harmonic representation
of the metric fluctuations, e.g. $A(\eta,x_i) = A(\eta,k) Y$.

The quantity $\delta w_Q$ represents spatial fluctuations in the equation of state and
specifies the pressure fluctuation as
\begin{equation}
\delta p_Q = (w_Q \delta_Q + \delta w_Q) \rho_Q \,.
\label{eqn:pressurefluctuation}
\end{equation}
Equation of state fluctuations
can arise from  intrinsic internal degrees of freedom in the dark energy or from
 temporal variations in the background equation of state through a general
coordinate or gauge transformation. 

The gauge transformation is defined as a perturbation in the
temporal and spatial coordinates of $x^\mu = \tilde x^\mu + (T,LY^{i})$  with respect to the
background under which \cite{Bar80}
\begin{align}
 A &=\tilde A - \yell{\dot T} - {\dot a \over a} \yell{T}\,, \nonumber\\
 B &=  \tilde B + \yell{\dot L} + k\yell{T} \,, \nonumber\\
 H_L &=  \tilde H_L - {k \over 3}\yell{L} - {\dot a \over a} \yell{T}\,, \nonumber\\
 H_T &=  \tilde H_T + k\yell{L}\,, 
\label{eqn:metrictrans}
\end{align}
for the metric and 
\begin{align} 
{\delta \rho_Q} &= {\delta\tilde\rho}_Q - \dot\rho_Q \yell{T}, \nonumber\\ 
{\delta  p_Q} &= {\delta \tilde p}_Q -\dot p_Q \yell{T}, \nonumber\\
 u_Q &=  \tilde u_Q - k(1+w_Q) \yell{T} \,,
\label{eqn:fluidtrans}
\end{align}
for the dark energy variables.  

Starting from the uniform density gauge
where there is no dark energy density fluctuation
$\delta \tilde\rho_{Q}=0$, there arises a contribution to the pressure fluctuation under
the gauge transformation [Eqn.~(\ref{eqn:fluidtrans})] of
\begin{align}
{\delta  p}_{Q}-\delta \tilde p_{Q} ={ \delta \rho_{Q} } c_{a}^{2}\,,
\label{eqn:gaugepressure}
\end{align}
where the adiabatic sound speed is
\begin{align}
c_{a}^{2} = &{\dot p_{Q}\over \dot \rho_{Q}} = { d p_Q/d \ln a \over d\rho_Q /d\ln a} \nonumber\\
                 = & w_Q - {1\over 3} {d \ln (1+w_Q) \over d\ln a}  \,.
                 \label{eqn:adiabaticsound}
\end{align}
If in the uniform density gauge the pressure fluctuation also vanishes  ($\delta \tilde p_{Q}=0$), 
the
dark energy only has adiabatic pressure fluctuations and from 
Eqns.~(\ref{eqn:pressurefluctuation}) and (\ref{eqn:gaugepressure})
\begin{align}
\delta w_{Q\, \rm adi}= & (c_{a}^{2}-w_Q)\delta_{Q} \,.
\end{align}
In this case, the pressure fluctuation and gravitational terms in the momentum conservation equation 
(\ref{eqn:perturbedgeneral}) become
\begin{align}
k c_a^2 \delta_Q + (1+w_Q)k A \,.
\label{eqn:pressuregravity}
\end{align}
If the dark energy accelerates the expansion $w_{Q} < -1/3$.
If $w_{Q}$ is also slowly varying,
 the adiabatic sound speed is imaginary: $c_{a}^{2} \sim w_Q < 0$.  
 In the small scale Newtonian approximation (and gauge), 
 the Poisson equation gives
 $ A \sim -\delta_Q / (k\eta)^2$ during the dark energy dominated epoch. 
 Hence for $c_a^2 < 0$ 
 pressure gradients would enhance potential gradients in 
 generating momentum density and  the
dark energy would collapse faster than the dark matter.
Thus for gravitational instability in the dark energy to be stabilized
by pressure gradients a source of non-adiabatic
pressure fluctuations is required.   These can be supplied by internal degrees
of freedom in the dark energy.  

Following \cite{Hu98}, the role of the background equation of state parameter as 
a closure relation between the pressure and density can be generalized
for an inhomogeneous dark energy component.   If one requires that
the pressure fluctuation is linear in the energy and momentum density fluctuations,
general covariance [or gauge invariance for linear fluctuations, see Eqn.~(\ref{eqn:fluidtrans})] requires that
\begin{align}
\delta w_Q |_{\rm non-adi} \propto \delta_Q + 3{\dot a \over a}
 {u_Q \over k} \,.
 \end{align}
We choose to specify the proportionality through an effective sound speed
such that the full equation of state fluctuation is given as  
\begin{align}
 \delta w_Q = & (c_{a}^{2}-w_Q)\delta_{Q} + (c_e^2 - c_a^2) \left(\delta_Q + 3{\dot a \over a}
 {u_Q \over k} \right) \,.
 \label{eqn:soundspeeddef}
 \end{align}
Note that $c_e^2=c_a^2$ for adiabatic pressure fluctuations.  The quantity
$c_e^2$ may in principle be a function of time and $k$.  

With this ansatz for $\delta w_{Q}$,
the covariant
 conservation equations (\ref{eqn:perturbedgeneral})  become
 \begin{align}
 \dot \delta_Q =& -3 {\dot a \over a} (c_e^2-w_Q) \delta_Q - 9 \left({\dot a \over a}\right)^2
 (c_e^2 - c_a^2) {u_Q \over k} \nonumber\\
 & - k u_Q   - (1+w_Q)(k B + 3 \dot H_L)\,, \nonumber\\
 \dot u_Q = & {\dot a \over a} [3(w_Q+c_e^2-c_a^2)-1]u_Q  + k c_e^2 \delta_Q \nonumber\\
 &- {2 \over 3} w_Q  (1-3K/k^2) k \pi_Q + (1+w_Q) k A \,.
 \label{eqn:perturbedconservation}
 \end{align}
Note that unlike in Eqn.~(\ref{eqn:pressuregravity}), 
it is $c_e^2$ and not $c_a^2$ that is the coefficient for $\delta_Q$ in the
momentum conservation equation; hence $c_e^2$ controls the stability of density 
fluctuations.  More formally in the zero momentum gauge or rest frame of the dark energy
where $u_Q=0$, Eqns.~(\ref{eqn:pressurefluctuation}) and (\ref{eqn:soundspeeddef}) imply
\begin{align}
\delta w_{Q\, \rm rest}  = (c_e^2 - w_Q) \delta_{Q \rm rest}\,, \nonumber\\ 
\delta p_{Q \, \rm rest}  = c_e^2  \delta \rho_{Q \rm rest}\,,
\end{align}
so that $c_e^2$ is the sound speed in the rest frame.  This property is useful
in the calculation of $c_e^2$ for a given dark energy model.

 Likewise, a closure relation between the anisotropic stress $\pi_Q$ and the
 energy and momentum density can also be supplied through a generalized
 equation of state \cite{Hu98}.  However unlike the non-adiabatic stress, a non-vanishing
 value is not required for stability by $w_Q < -1/3$.  
 We shall see now that for scalar field dark energy candidates
 it in fact does vanish in linear theory.  

\subsection{Quintessence and k-Essence Fields}

The stress-energy
tensor for a scalar field $Q$ with a Lagrangian $F(X,Q)$ is \cite{GarMuk99}
\begin{align}
\yell{T^\mu_{\hphantom{0}\nu}} =  {\partial F\over \partial X}\nabla^\mu Q \nabla_\nu Q   + F \delta^\mu_{\hphantom{i}\nu} \,,
\label{eqn:scalarfieldt}
\end{align}
where the kinetic term
\begin{align}
X = -{1 \over 2} \nabla^\mu  Q \nabla_\mu Q \,.
\end{align}
For the case of a canonical kinetic energy term 
\begin{align}
F(X,Q) =  X -  V_Q \,,
\label{eqn:canonical}
\end{align}
where $V_Q(Q)$ is the scalar field potential. $Q$ is then called the quintessence field.
For the case $F(X,Q) = -X-V_Q$, the dark energy is called a phantom field (e.g. \cite{Cal02})
and for a more general modification of the kinetic term, a k-essence field \cite{ArmMukSte00}.

Matching terms between Eqn.~(\ref{eqn:stressenergy}) and (\ref{eqn:scalarfieldt}),
and dropping contributions that are second order in the spatial gradients of the field
\begin{align}
\rho_Q & = 2 {\partial F\over \partial X} X  - F\,, \nonumber\\
p_Q & = F\,,\nonumber\\
q_{Q\, i } & = -{\partial F \over \partial X} a^{-2} \dot Q \nabla_i Q \,, \nonumber\\
 {\Pi_{Q}} {}^{i}_{\hphantom{i}j} &= 0  \,.
 \label{eqn:backgroundfield}
 \end{align}
 In particular the density and pressure fluctuations become
 \begin{align}
 \delta \rho_Q = &(2 {\partial^2 F \over \partial X^2} X + {\partial F \over \partial X})\delta X 
 + \left(  2 {\partial^2 F \over \partial Q \partial X} X  - {\partial F \over \partial Q} \right) \delta Q\,,\nonumber
\\
\delta p_Q = & {\partial F \over \partial X} \delta X  + {\partial F \over \partial Q} \delta Q\,,
\label{eqn:perturbedfield}
 \end{align}
 where 
 \begin{equation}
 \delta X = a^{-2} (\dot Q \dot {\delta Q} - A \dot Q^2)  \,.
 \end{equation}
 
 Note that under a gauge transformation $Q$ transforms as a scalar
 \begin{align}
 {\delta Q} &= {\delta\tilde Q} - \dot Q \yell{T}\,,
  \end{align}
  and
  \begin{align}
  \rho_Q u_Q = {\partial F \over \partial X} a^{-2} \dot Q k \delta Q \,.
  \label{eqn:fieldmomentum}
  \end{align}
 Hence  the uniform field gauge where $\delta Q = 0$ coincides with the
 rest gauge $u_Q=0$.   Since this is the gauge in which $c_e^2 = \delta p_Q/\delta \rho_Q$,
 the effective sound speed of a scalar field is
 \cite{GarMuk99},
 \begin{align}
 c_e^2 = { \partial F /\partial X \over 2 (\partial^2 F /\partial X^2) X + (\partial F/\partial X) }\,.
 \label{eqn:kessencesound}
 \end{align}
 For a scalar field with the canonical kinetic term in Eqn.~(\ref{eqn:canonical}), $c_e^2=1$.   
 More generally, to achieve a constant $c_e^2$ one requires
 \begin{equation}
 {\partial F \over \partial X} = \left( {X \over V_K} \right)^{1-c_e^2 \over 2 c_e^2} \,,
 \label{eqn:dfdx}
 \end{equation}
 where $V_K$ is a constant with dimensions of energy density.
 If for simplicity one further assumes that the kinetic and potential terms are additive
$\partial F /  \partial Q =-{\partial V_Q/ \partial Q}$ and
\begin{equation}
F= {2 c_e^2 \over 1+ c_e^2} \left( { X \over V_k} \right)^{1+c_e^2 \over 2 c_e^2}V_k  - V_Q \,.
\label{eqn:constantce2}
\end{equation} 

The effective sound speed of Eqn.~(\ref{eqn:kessencesound}) along 
 with $w_Q$ in the background closes the evolution equations
 Eqn.~(\ref{eqn:perturbedconservation}) once the metric fluctuations are 
 specified by the Einstein equations (see \S \ref{sec:initialcond}).
 The evolution of dark energy density perturbations is then completely defined by a choice of
initial conditions for $\delta_Q$ and $u_Q$.
 
 \subsection{Background Evolution}

To obtain the background evolution of the dark energy, one begins with the
 energy density conservation
equation (\ref{eqn:backgroundrho}) in terms of the field variables
\begin{align}
\ddot Q + (3c_e^2 -1) {\dot a \over a} \dot Q + c_e^2 a^2 \left( {\partial F \over \partial X} \right)^{-1}
 {\partial \rho_Q \over \partial Q}
 = 0
\end{align}
or
\begin{align}
{d^2 Q\over d\ln a^2} + {3\over 2}(2 c_e^2 - 1 - w_T) {d Q \over d\ln a}   +  {c_e^2 \over H^2} \left( {\partial F \over \partial X} \right)^{-1}
 {\partial \rho_Q \over \partial Q} = 0  \,. 
 \label{eqn:Qeqnmotion}
\end{align}
For the canonical kinetic term,
\begin{equation}
{d^2 Q\over d\ln a^2}  + {3 \over 2}(1-w_T) {d Q \over d\ln a}+ {{1 \over H^2}{d V_Q \over dQ}} = 0\,,
\label{eqn:KleinGordon}
\end{equation}
where $w_T = p_T /\rho_T$ is the equation of state for
the sum of all components.   Note that $w_T$ depends on $Q$ and hence in practice
it is more convenient
to solve a set of coupled first order differential equations in $Q$ and $Hd Q/d\ln a = \sqrt{2 X}$.
Likewise, since $1+w_Q $ appears in the evolution equations, one calculates this quantity directly
as 
\begin{align}
1+w_Q = {2 (\partial F/\partial X) X \over 2 (\partial F /\partial X)X - F } \,.
\end{align}
Combined with Eqn.~(\ref{eqn:dfdx}), this relation implies that 
if the field is potential energy dominated $w_Q \rightarrow -1$ and if it is kinetic energy
dominated $w_Q \rightarrow c_e^2$. 

With equation (\ref{eqn:Qeqnmotion}), the expression for the adiabatic sound speed 
becomes
\begin{equation}
c_a^2 = c_e^2 + {c_e^2 \partial \rho_Q / \partial Q - \partial F/\partial Q
\over 
3 H^2 (\partial F/ \partial X) ({d Q / d\ln a})} \,.
 \end{equation}
 The adiabatic sound speed takes on a specific form in the case where the Hubble
 drag inhibits the motion of the field.  In this case $\partial \rho_Q / \partial Q$ is nearly constant.  
 If $w_T$ is constant, $dQ / d\ln a$ reaches a ``terminal velocity" that is independent of the initial conditions and scales with $H$.  
 Utilizing the constant effective sound speed
 form for $F$ in equation~(\ref{eqn:constantce2}), the equation of motion (\ref{eqn:Qeqnmotion})
 can be written in the form of the Bernoulli equation and the adiabatic sound speed becomes
 \begin{equation}
 c_a^2 = c_e^2 -{1 \over 2}(c_e^2 +1 )(3+w_T)\,.
 \label{eqn:terminal}
 \end{equation}
 For the case of a canonical kinetic term $c_a^2 = -2 - w_T$ \cite{BarCorLidMal04} (see also
 Fig.~\ref{fig:qevol}).
 
 On the other hand, if the Hubble drag is negligible 
 \begin{equation}
 {dQ \over d \ln a} \propto a^{-3(2c_{e}^{2}-1-w_{T})/2} \,,
 \end{equation}
 and assuming kinetic energy domination $w_{Q}=c_{a}^{2}=c_{e}^{2}$.
 
\subsection{Initial Conditions and Evolution}
\label{sec:initialcond}
We numerically solve the dark energy evolution equations in comoving gauge where
the total momentum density vanishes
\begin{align}
\rho_T u_T \equiv \sum_i \rho_i u_i =0\,.
\end{align}
Here  $i$ runs over all species of energy density.  We define the total density and pressure
fluctuation similarly.  The auxiliary condition 
$H_T=0$ completely fixes the coordinate freedom.
We call the remaining metric
degrees of freedom  $\zeta \equiv H_L$, $\xi \equiv A$ and $V_T \equiv B$, where 
$V_T$ also has the interpretation of the total momentum weighted velocity [see \cite{Bar80} and
the discussion below
Eqn.~(\ref{eqn:fluidfluct})].
The Einstein equations become
 \begin{align}
 \dot \zeta + {K \over k} V_T =& {\dot a \over a}\xi \nonumber \\
=& \dot{a \over a} {1 \over \rho_T+p_T} \left[ - \delta p_T
  +
 {2 \over 3}\left( 1- {3K\over k^2}\right) p_T \pi_T \right]\,, \nonumber\\
 V_T =& (\Phi - \zeta) k \left( { \dot a \over a} \right)^{-1}
 \label{eqn:bardeencurvature}
 \end{align}
 where recall $K$ is the background spatial curvature.   
 The continuity equation for the total density perturbation completes
 the basic equations
 \begin{align}
 \dot \delta_T = -3 {\dot a \over a} \delta w_T - (1+w_T) (k V_T + 3 \dot \zeta) \,,
 \label{eqn:continuitytotal}
 \end{align}
 as  the momentum conservation equation was used to relate $\xi$ to the
 stresses in Eqn.~(\ref{eqn:bardeencurvature}).  
 In that equation
 \begin{align}
 (k^2 - 3K)\Phi = 4\pi G a^2 \rho_T \delta_T
 \label{eqn:phi}
 \end{align}
 and is equal to the curvature fluctuation in the Newtonian or longitudinal gauge 
 $\Phi \equiv H_L |_{\rm Newt}$.  For reference, the Newtonian potential $\Psi \equiv A|_{\rm Newt}$ employed in
 \S \ref{sec:transfer} 
 is
 \begin{align}
 \Psi = -\Phi - 8\pi G a^2 p_T\pi_T/k^2 \,,
 \label{eqn:psi}
 \end{align}
 and the total velocity obeys an auxiliary relation
 \begin{align}
 \dot V_T + {\dot a \over a} V_T= - k (\xi - \Psi)
 \end{align}
 that is useful for checking numerical solutions.   
 
Equation (\ref{eqn:bardeencurvature}) implies that
the curvature fluctuation $\zeta$ changes only in response to stress fluctuations.
As in the discussion of dark energy perturbations above, the evolution equations
are closed through an assumption for the stress perturbations.    In practice, for numerical
solutions that extend to subhorizon scales,
we replace Eqn.~(\ref{eqn:continuitytotal}) with the conservation equations for
the individual energy density species to implicitly solve for $\delta w_T$.
   
For the initial conditions, assuming nearly adiabatic stresses
\begin{equation}
\left| \delta p_T -{\dot p_T \over \dot \rho_T}\delta\rho_T \right| \ll | \delta p_T |
\label{eqn:adiabaticic}
\end{equation}
and a closure relation for the anisotropic stress of the neutrinos which
comes from the Boltzmann equation [e.g. \cite{Hu98} Eqn. (12)]
\begin{align}
\pi_\nu = {4 \over 5} k\eta V_T \,,
\end{align}
we obtain the solution to the evolution equations in the radiation dominated regime
\begin{align}
\delta_T =& {4 \over 9} {1 + 3\alpha_\nu \over 1+ 2 \alpha_\nu} (1-3{K\over k^2}) (k\eta)^2 \zeta_i \,, \nonumber\\
V_T =& -{1\over 3} {1\over 1+ 2 \alpha_\nu} (k\eta) \zeta_i \,,
\label{eqn:denvel}
\end{align}
where $\alpha_\nu = 2 \rho_\nu / 15\rho_r$ accounts for the
anisotropic stress of the neutrinos
\begin{align}
\pi_T  = &{\rho_\nu \over \rho_r}\pi_\nu \nonumber\\
           =  & -{2\alpha_\nu \over 1 + 2\alpha_\nu} (k\eta)^2\zeta_i \,.
\end{align}
Here $\rho_r = \rho_\gamma+\rho_\nu$, the total radiation density and we have kept only
leading order terms in power of $k\eta$.
Given that the stress perturbation $\delta p_T /(\rho_T+p_T) = {\cal O}[(k\eta)^2] \zeta_i$, 
Eqn.~(\ref{eqn:bardeencurvature}) shows that
the curvature perturbation is nearly constant for superhorizon
 adiabatic stresses $\zeta(k,\eta_i) = \zeta_i(k)$ \cite{Bar80}.

  
The dark energy perturbations associated with the curvature fluctuation can then
be related to the total density perturbation by substituting these relations back into
the Einstein equations (\ref{eqn:bardeencurvature}) for $\dot \zeta$ and $\xi$
and employing them in the conservation equations (\ref{eqn:perturbedconservation}).
The result is
\begin{align}
\delta_Q &= A_\delta (1+w_Q) \delta_T \,,\nonumber\\
u_Q         &= A_u (1+w_Q) (k\eta) {\delta_T} \,,
\label{eqn:adiabatic}
\end{align}
where assuming a radiation dominated initial condition
\begin{align}
A_u & = {3 c_e^2 + (1+6\alpha_\nu)(3w_Q-2) \over  4(1+3\alpha_\nu)
[8 + 3c_e^2(2-3c_a^2+  3 w_Q) -12 w_Q] }\,,\nonumber\\
A_\delta & = {3 \over 4} { 1- 6(c_e^2 - c_a^2) A_u \over 1- 3(w_Q - c_e^2)/2}\,.
\end{align}
We have here assumed that $w_Q$ and $c_a^2$ are nearly constant compared
with the expansion time but have allowed $(1+w_Q)$ to vary as appropriate
for a scalar field under the Hubble drag by factoring this quantity out of 
Eqn.~(\ref{eqn:adiabatic}).

The curvature initial conditions for the dark energy
take on this rather intricate form involving $A_u$ in the
relation for $A_\delta$ since the internal non-adiabatic stress
of the dark energy cannot be neglected even though its contribution to the
total non-adiabatic stress in Eqn.~(\ref{eqn:adiabaticic}) can be ignored.  In general,
the curvature mode always carries non-adiabatic stresses beyond the leading order in $(k\eta)$
and $\rho_i/\rho_r$ where $i \ne \gamma, \nu$.  
Formally these vanish if the initial conditions are taken to $\eta_i \rightarrow 0$.

Deviations in the initial conditions for the dark energy from these relations represents an isocurvature
mode since the dark energy is assumed to carry a negligible fraction of the net
energy density at the initial conditions so that the radiation density fluctuation
$\delta_{r}=\delta_{T}$.  If we also assume that the relative deviations are large
$\delta_{Q}(\eta_i,k) \gg (1+w_Q) \delta_T(\eta_i,k)$ then the dark energy-radiation
entropy fluctuation 
\begin{align}
(1+w_Q) S_{Qr} \equiv & \delta_Q - {1+w_Q \over 1+ w_r} \delta_r \,, \nonumber\\
\approx & \delta_Q  \,,
\end{align}
which corresponds to the usual definition of the isocurvature mode 
as being generated by $S_{Qr}$.
The distinction here is that the adiabatic mode has $S_{Qr} ={\cal O}(\delta_T) \ne 0$
due to the intrinsic entropy of $Q$ and evolution of $w_Q$.  

Neglecting the metric fluctuations generated by the dark energy fluctuations, we find
that the  
the evolution equations (\ref{eqn:perturbedconservation})
are solved by a linear combination of the adiabatic mode
and
\begin{align}
\delta_Q  &  = I_\delta (k\eta)^p \,, \nonumber\\
u_Q          & = I_u(k\eta)^{p+1} \,,
\end{align}
where
\begin{align}
{I_\delta \over I_u} =& \left[ p + {3 (1+w_T + 2 c_a^2 - 2 c_e^2 - 2 w_Q) \over 1+ 3 w_T} \right] c_e^{-2}  \nonumber\\
\end{align}
and $p$ solves the equation
\begin{align}
\left[ p + {6 \over 1+3 w_T}(c_e^2-w_Q) \right]  {I_\delta \over I_u}= -9\left( {2 \over 1+3 w_T} \right)^2 (c_e^2-c_a^2) \,.
\end{align}
For initial conditions in the radiation dominated era, the solutions to this quadratic
equation are
\begin{align}
p = {-1 -{3\over 2}c_a^2}+ 3w_Q \pm \sqrt{(1+3 c_a^2/2)^2 - 6 c_e^2}\,.
\end{align}
These solutions in fact apply for essentially all gauges 
until the epoch of dark energy domination.   The only exceptions are those that
place explicit conditions on the dark energy fluctuations such as $\delta_Q=0$
or $u_Q=0$.

It is instructive to consider the two limiting cases of Hubble drag domination 
 and negligible Hubble drag or kinetic energy
domination in Eqn.~(\ref{eqn:terminal}).    In the former case,
\begin{align}
p= 0,  \quad {3 \over {1 + 3 w_T}} (-2 + c_e^2 + c_e^2 w_T) \,.
\end{align}
In terms of the field variables, the constant mode corresponds to an initial condition
where $\delta X=0$, i.e. the kinetic energy terms in Eqn.~(\ref{eqn:perturbedfield}) 
for the density perturbation vanish.   Therefore the potential energy fluctuation 
remains constant. Note that in spite of this the momentum density
does not vanish.
The second mode corresponds
to an initial conditions with comparable kinetic and potential energy ($\delta Q$ terms) 
in the  perturbation.
An arbitrary initial condition can be decomposed into a superposition of the modes.
Note that the second mode is decaying for $c_e^2 < 2/(1+w_T)$ and growing for $c_e^2 > 2/(1+w_T)$.
For a canonical kinetic term, this is a decaying mode for all $w_T < 1$.
Nonetheless,  if $2 > c_e^2 > 3/2$ a small
initial density or field fluctuation will be amplified during radiation domination and
freeze in during matter domination even though the background field is potential energy dominated
during the whole expansion history $w_Q \approx -1$.   

The existence 
of this mode could potentially allow a solution to the 
gravitational wave problem of \S \ref{sec:inflation}
 by amplifying the field fluctuations from their initial conditions. 
However the amount of amplification is dependent on the initial ratio of kinetic
to potential energy in the fluctuation. It also depends on the ratio in the 
background 
since the fractional
fluctuation in the kinetic energy density (as opposed to the total energy density)
 must also remain small for the mode analysis  to remain valid.  
Finding an explicit model that satisfies these conditions is beyond
the scope of this work. 

In the opposite regime of kinetic energy domination, the two solutions become
\begin{align}
p = 0 , \quad { {3 (2c_e^2 - 1 - w_T) \over 1+ 3w_T}} \,.
\label{eqn:kineticdominated}
\end{align}
The $p=0$ solution corresponds to a kinetic energy dominated perturbation where
$\delta X$ scales with $X$.  The other solution corresponds to a pure velocity
isocurvature mode where $u_Q/(k\eta) \gg \delta_Q \approx 0$.
Note that $u_Q$ grows if $c_e^2>1/3$ as is the
case for the canonical kinetic term.    In the field representation, this mode represents 
a case where the density perturbation is dominated by the potential energy
and hence negligible in the kinetic energy dominated regime.  The
field fluctuation $\delta Q$ then remains constant but the momentum density $u_Q$ in
Eqn.~(\ref{eqn:fieldmomentum}) grows due to the redshifting of $\rho_Q$ in the background.
If the field later exits from kinetic energy domination, the field fluctuation then becomes
an energy density fluctuation.

Finally, there is the well-studied tracking regime where $1+w_Q \propto 1+ w_T$.
Then $1+w_Q$ is approximately constant and Eqn.~(\ref{eqn:adiabaticsound}) gives
$c_a^2 \approx w_Q$.  The two solutions
become
\begin{align}
p = &-{3 \over 2(1+3w_T)}
\Big( 1- 2w_Q + w_T  \nonumber\\
&  \pm \sqrt{(1+2w_Q + w_T)^2 -8 c_e^2(1+w_T) } \Big) \,.
\end{align}
The index $p$ is maximized by maximizing $w_Q$.  The maximum $w_Q = c_e^2$ for 
a kinetic energy dominated field and hence the fastest growing modes are given by Eqn.~(\ref{eqn:kineticdominated}).
By further requiring the tracker condition 
$w_Q < (1+w_T)/2$ \cite{SteWanZla99},
we see that Re$(p) < 0$ and 
there are again no growing modes.
The field fluctuations then ``track" and lose
their dependence on the initial isocurvature perturbation as is well known.    Note that for the canonical
kinetic term in the tracking regime, the proportionality is \cite{SteWanZla99}
\begin{equation}
1+w_Q \approx {1 + w_T \over 2\Gamma-1}
\end{equation}
where $\Gamma$ was defined in Eqn.~(\ref{eqn:trackercond}). The perfect tracker
is attained at the limiting case of $1+w_Q=1+w_T$ or $\Gamma=1$
which is acheived for an purely exponential potential
$V_Q \propto e^{-C Q}$ where $C$ is constant  \cite{FerJoy97}.
 Here the dark energy remains
a constant fraction of the total energy density.
\vfill


\end{document}